\documentclass[preprint,eqsecnum,aps,nofootinbib]{revtex4}
\usepackage{amsfonts,amsmath,amssymb,amsthm}
\usepackage{latexsym}
\usepackage{bbm,bm}
\usepackage{graphicx}
\usepackage{tikz} 

\newcommand{\ket}[1]{\lvert #1 \rangle}
\newcommand{\bra}[1]{\langle #1 \lvert}
\newcommand{\beq}{\begin{equation}}
\newcommand{\eeq}{\end{equation}}
\newcommand{\beqs}{\begin{eqnarray}}
\newcommand{\eeqs}{\end{eqnarray}}
\begin{document}

\title{Tripartite entanglement and matrix inversion quantum algorithm }

\author{Mi-Ra Hwang$^1$, MuSeong Kim$^1$,  Eylee Jung$^1$, Chang-Yong Woo$^1$, and DaeKil Park$^{1,2}$\footnote{dkpark@kyungnam.ac.kr} }

\affiliation{$^1$Department of Electronic Engineering, Kyungnam University, Changwon
                 631-701, Korea    \\
             $^2$Department of Physics, Kyungnam University, Changwon
                  631-701, Korea    
                      }

\begin{abstract}
The role of entanglement is discussed in the Harrow-Hassidim-Lloyd (HHL) algorithm. We compute all tripartite entanglement at every steps of the HHL algorithm.
The tripartite entanglement is generated in the first quantum phase estimation (QPE) step. 
However, it turns out that amount of the generated entanglement is not maximal except very rare cases. In the second rotation step some tripartite entanglement is 
annihilated. Thus, the net tripartite entanglement is diminished. At the final inverse-QPE step the matrix inversion task is completed at the price of complete 
annihilation of the entanglement. An implication of this result is discussed.

\end{abstract}

\maketitle

\section{Introduction}

In quantum information processing quantum entanglement\cite{text,schrodinger-35,horodecki09} plays an important role as a physical resource. 
It is used in various quantum information processing, such as  quantum teleportation\cite{teleportation,Luo2019},
superdense coding\cite{superdense}, quantum cloning\cite{clon}, quantum cryptography\cite{cryptography,cryptography2}, quantum
metrology\cite{metro17}, and quantum computer\cite{qcreview,computer,supremacy-1}. 
Furthermore, quantum computers with hundred qubits were already constructed in IBM and Google. 
In this reason the development of quantum algorithms becomes important more and more to use the quantum computers efficiently.

The representative of the quantum algorithm are Shor' factoring\cite{shor94} and Grover's search\cite{grover96} algorithms.
Few years ago another quantum algorithm\cite{hhl} called
Harrow-Hassidim-Lloyd (HHL) algorithm was developed. It is a quantum algorithm, which can compute the inverse of sparse matrix. 
If $A$ is a $N \times N$ sparse matrix, HHL algorithm completes the inversion task with a runtime of ${\cal O} \left( s^2 \kappa^2 \log(N) / \epsilon \right)$, where
$s$ is the maximum number of non-zero entries in  row or column, $\kappa$ the condition number, and $\epsilon$ the precision. 
Since same task can be implemented in the classical computer with a runtime of ${\cal O} \left( s \kappa N \log(1 / \epsilon) \right)$ even though  the most efficient algorithm is adopted, 
one can say that the HHL algorithm improves exponentially in the matrix  inversion task over the best classical algorithm.
This algorithm is based on the efficient Hamiltonian simulation\cite{berry05}. Subsequently, there was a proposals to deal with dense matrices\cite{dense}. Also a hybrid algorithm\cite{hybrid} was proposed, where 
both classical and quantum computers are used appropriately. 

In this paper we examine a question:  `how much the HHL algorithm uses the quantum entanglement efficiently?'. 
In order to explore this issue we compute the tripartite entanglement in every steps of the HHL algorithm by choosing an experimental realization of  the algorithm presented in Ref. \cite{exp-hhl-1}.
Since the HHL algorithm is known  to be optimal\cite{hhl} in the matrix inversion task and entanglement is known as a physical resource, we expect that much entanglement is created and finally annihilated during the process of the algorithm.
However, it turns out that maximal tripartite entanglement is not created except very rare cases. As expected, all created entanglement is annihilated at the final step of the algorithm.

This paper is organized as follows. In Sec. II we discuss the role of entanglement in the Grover's algorithm. In Sec. III we compute the three-tangle\cite{ckw} of some rank-$2$ mixture. The result of this section will be 
used in Sec. V. In Sec. IV we briefly review the HHL algorithm. In Sec. V we compute the tripartite entanglement at every steps of the HHL algorithm. In Sec. VI a conclusion and further discussion  are presented. 
In appendix A  the quantum phase estimation (QPE) is discussed when the unitary operator $U$ is operated to the non-eigenvector $\ket{\psi}$. The result of this appendix is used in Sec. IV.

\section{Entanglement in Grover's algorithm}
%%%%%%%%%%%%%%%%%%%%%%%%%%%%%%%%%%%%%%%%%%%%%%%%%%%%%%%%%
\begin{figure}[ht!]
\begin{center}
\includegraphics[height=6.0cm]{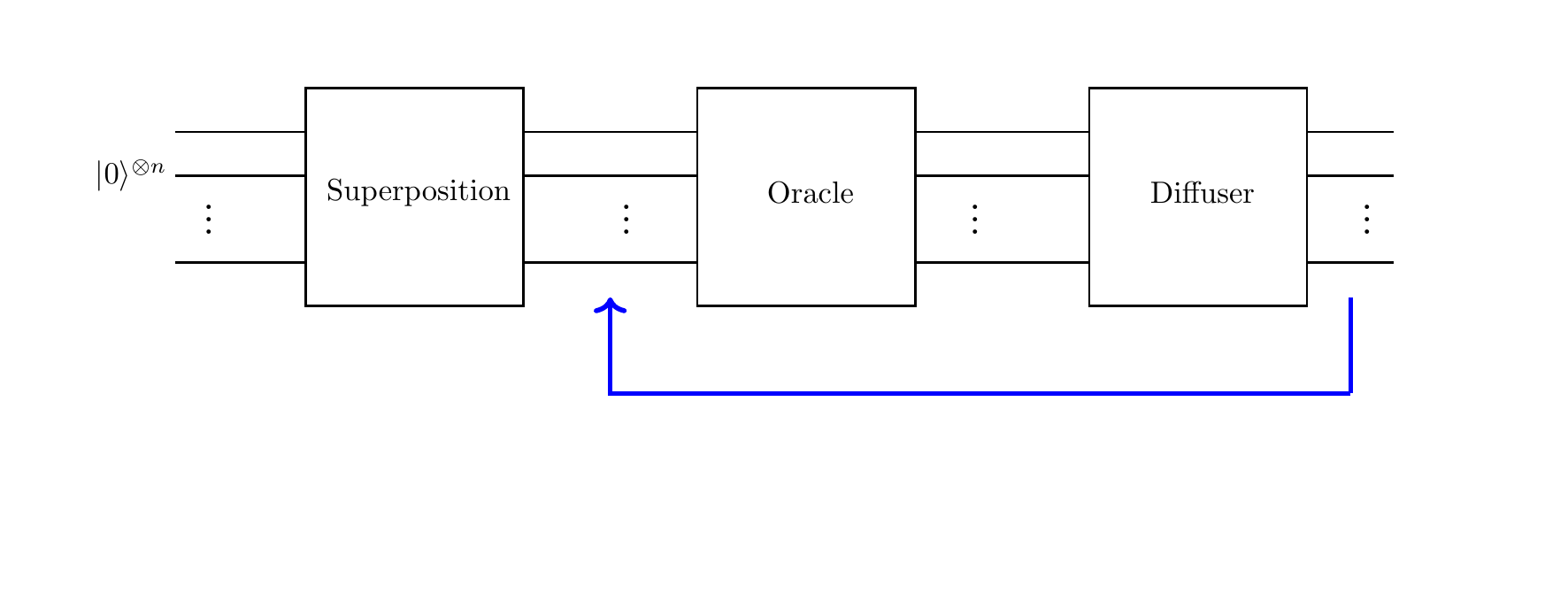} 

\caption[fig1]{(Color online) The schematic representation of the Grover's algorithm. The oracle and diffuser unitary transformations should be repeated roughly $\sqrt{N}$ times for large $N$.
 }
\end{center}
\end{figure}
%%%%%%%%%%%%%%%%%%%%%%%%%%%%%%%%%%%%%%%%%%%%%%%%%%%%%%%%%%%

Let us consider a set ${\cal S} = \left\{ \ket{j} | j = 0, 1, \cdots ,N-1 \right\}$ with $\bra{j_1} j_2 \rangle = \delta_{j_1 j_2}$.
Grover's algorithm\cite{grover96,grover97} is a quantum algorithm, which tries to find a particular quantum state $\ket{\psi_G} \in {\cal S}$. 
Grover's algorithm consists of three unitary transformations called superposition, oracle, and diffuser. The superposition transforms the initial state $\ket{0}^{\otimes n}$ to 
a superposed state, where all states in ${\cal S}$ are superposed with equal probability amplitude. This can be achieved by making use of the Hadamard gate
\begin{eqnarray}
\label{hadamard}
H = \frac{1}{\sqrt{2}} \left(     \begin{array}{cc}
                                                1  &  1                   \\
                                                1  &  -1
                                                \end{array}                           \right).
\end{eqnarray}
Therefore, after superposition transformation the initial state is changed into 
\begin{equation}
\label{post-superposition}
\ket{s} = H^{\otimes  n} \ket{0}^{\otimes n} = \frac{1}{\sqrt{N}} \sum_{j=0}^{N-1} \ket{j}
\end{equation}
where $N = 2^n$. The oracle and diffuser are described by the unitary operators $U_{\mbox{oracle}} = \openone - 2 \ket{\psi_G} \bra{\psi_G}$ and $U_{\mbox{diffuser}} = 2 \ket{s} \bra{s} - \openone$, respectively. 
The oracle changes a sign of $\ket{\psi_G}$ in $\ket{s}$. The diffuser increases the probability amplitude of $\ket{\psi_G}$ from $U_{\mbox{oracle}} \ket{s}$. Given $N$ the oracle and diffuser transformations are repeated roughly 
$\sqrt{N}$ times\cite{bennett97,boyer96,grover98}\footnote{As proved in Ref. \cite{grover98}, the optimal number of queries is $\pi \sqrt{N} / 4$ when $N$ is very large.} for large $N$.
The schematic quantum circuit of the Grover's algorithm is plotted in Fig. 1.

The role of entanglement manifestly appears when $N = 4$. If we assume $\ket{\psi_G} = \ket{3} = \ket{11}$ in this case, $U_{\mbox{oracle}}$ transforms $\ket{s}$ into
\begin{equation}
\label{oracle-1}
\ket{\psi_{\mbox{oracle}}} = U_{\mbox{oracle}} \ket{s} =\frac{1}{2} \left( \ket{00} + \ket{01} + \ket{10} - \ket{11} \right).
\end{equation}
Since the concurrence, one of the bipartite entanglement measure, is defined as ${\cal C} = 2 |a_{00} a_{11} - a_{01} a_{10}|$ for two qubit pure state $\ket{\psi} = \sum_{i,j=0}^1 a_{ij} \ket{ij}$ \cite{form2,form3}, 
$U_{\mbox{oracle}}$ changes the separable state $\ket{s}$ into the maximally entangled state $\ket{\psi_{\mbox{oracle}}}$. Then, $U_{\mbox{diffuser}}$ exactly detects $\ket{\psi_G}$, which means 
$U_{\mbox{diffuser}} \ket{\psi_{\mbox{oracle}}} = \ket{\psi_G}$.
Therefore, $U_{\mbox{oracle}}$ creates an entanglement maximally and $U_{\mbox{diffuser}}$ increases the probability amplitude of $\ket{\psi_G}$ maximally at the price of complete annihilation of entanglement. 

\begin{center}
\begin{tabular}{c|c|c|c|c} \hline \hline
                            \hspace{.1cm}  &\hspace{.1cm}  $\ket{\psi_1}$ \hspace{.1cm} & \hspace{.1cm}  $\ket{\psi_2}$ \hspace{.1cm} &  \hspace{.1cm} $\ket{\psi_3}$ \hspace{.1cm} &  \hspace{.1cm} $\ket{\psi_4}$                                                                 \\   \hline  
$\tau_3$              \hspace{.1cm}   & $1/4$    & $1 / 16$  &  $9/64$  & \hspace{.1cm}  $9/256$                                                                                                                       \\ 
${\cal C}_{AB}$   \hspace{.1cm}  &  $1/2$   &   $1/4$  &  $3/8$  & \hspace{.1cm} $3 / 16$                                                                                                                \\    
${\cal C}_{AC}$  \hspace{.1cm}  &  $1/2$   &   $1/4$  &  $3/8$  & \hspace{.1cm}  $3 / 16$                                                                                                                \\   
${\cal C}_{BC}$   \hspace{.1cm} &  $1/2$   &   $1/4$  &  $3/8$  & \hspace{.1cm} $3 / 16$                                                                                                                \\  \hline
\end{tabular}

\vspace{0.3cm}
Table I: Entanglement flow in Grover's algorithm when $N = 8$
\end{center}
\vspace{0.5cm}

Even though Grover's algorithm is optimal as a quantum searching algorithm\cite{zalka99}, such maximal creation and complete annihilation of entanglement do not occur for large $N$. 
For example, let us consider $N = 8$ case. If $\ket{\psi_G} = \ket{7} = \ket{111}$, Grover's algorithm changes the quantum state as 
\begin{eqnarray}
\label{N8grover}
&&\ket{\psi_1} = U_{\mbox{oracle}} \ket{s} = \frac{1}{2 \sqrt{2}} \left( \ket{0} + \ket{1} + \ket{2} + \ket{3} + \ket{4} + \ket{5} + \ket{6} - \ket{7}  \right)    \\   \nonumber
&&\ket{\psi_2} = U_{\mbox{diffuser}} \ket{\psi_1} = \frac{1}{4 \sqrt{2}} \left( \ket{0} + \ket{1} + \ket{2} + \ket{3} + \ket{4} + \ket{5} + \ket{6} + 5 \ket{7}  \right)    \\   \nonumber
&&\ket{\psi_3} = U_{\mbox{oracle}} \ket{\psi_2} = \frac{1}{4 \sqrt{2}} \left( \ket{0} + \ket{1} + \ket{2} + \ket{3} + \ket{4} + \ket{5} + \ket{6} - 5 \ket{7}  \right)    \\   \nonumber
&&\ket{\psi_4} = U_{\mbox{diffuser}} \ket{\psi_3} = - \frac{1}{8 \sqrt{2}} \left( \ket{0} + \ket{1} + \ket{2} + \ket{3} + \ket{4} + \ket{5} + \ket{6} - 11 \ket{7}  \right).
\end{eqnarray}
The three-tangles $\tau_3$\cite{ckw} for $\ket{\psi_i}$ and concurrences\cite{form2,form3} for the reduced mixed states are summarized in Table I. As Table I shows, $U_{\mbox{oracle}}$ transforms the 
separable state $\ket{s}$ to $\ket{\psi_1}$, whose three-tangle is $1/4$. Since maximum three-tangle is $1$\footnote{Maximal three-tangle is realized in the Greenberger-Horne-Zeilinger (GHZ) state\cite{dur00-1} 
$$\ket{GHZ} = \frac{1}{\sqrt{2}} (\ket{000} + \ket{111})$$ and its local unitary transformed states.}, $\ket{\psi_1}$ is only partially entangled state. Thus, maximal entanglement creation does not occur in this case. 
$U_{\mbox{diffuser}}$ changes $\ket{\psi_1}$ into $\ket{\psi_2}$, whose three-tangle is $1/16$. Therefore, the complete annihilation of tripartite entanglement also does not occur. Same is true for $\ket{\psi_3}$ and $\ket{\psi_4}$. 
Similar behavior occurs in the bipartite entanglement.

\section{three-tangle for rank-$2$ mixed state $\rho$}

In this section we compute the three-tangle of following rank-$2$ mixed state $\rho$ for later use. 
The state is given by 
\begin{equation}
\label{rank2-1}
\rho = p \ket{\phi_1} \bra{\phi_1} + (1 - p) \ket{\phi_2} \bra{\phi_2}
\end{equation}
where $0 \leq p \leq 1$ and 
\begin{eqnarray}
\label{rank2-2}
&&\ket{\phi_1} = \frac{x_1}{\sqrt{2}} \left( \ket{010} - \ket{011} \right) + \frac{x_2}{\sqrt{2}} \left( \ket{100} + \ket{101} \right)                  \\    \nonumber
&&\ket{\phi_2} = -\frac{x_2}{\sqrt{2}} \left( \ket{010} - \ket{011} \right) + \frac{x_1}{\sqrt{2}} \left( \ket{100} + \ket{101} \right)  
\end{eqnarray}
with $0 \leq x_1 \leq 1$ and $x_2 = \sqrt{1 - x_1^2} \geq 0$. It is straightforward to show that the three-tangles of $\ket{\phi_1}$ and $\ket{\phi_2}$ are the same as $4 x_1^2 x_2^2$. 
Therefore, the three-tangle of $\rho$ should satisfy
\begin{equation}
\label{rank2-3}
\tau_3 (\rho) = 4 x_1^2 x_2^2   \hspace{1.0cm} \mbox{when} \hspace{.2cm} p = 0 \hspace{.2cm} \mbox{or} \hspace{.2cm} p=1.
\end{equation}

For mixed states the three-tangle is defined by a convex-roof
method\cite{benn96,uhlmann99-1} as follows:
\begin{equation}
\label{mixed-3-tangle}
\tau_{ABC} (\rho) = \min \sum_i p_i \tau_{ABC} (\rho_i),
\end{equation}
where the minimum is taken over all possible ensembles of pure states. The pure state ensemble
corresponding to the minimum $\tau_{ABC}$ is called optimal decomposition. It is in general
difficult to derive the optimal decomposition for arbitrary mixed states.
The three-tangle for nontrivial rank-$2$ mixed state was explicitly computed in \cite{tangle2}. 
In Ref. \cite{convex_hull} it was shown that the three-tangle can be computed by constructing the convex hull in the minimum of characteristic curves.
Using such a method the three-tangles for higher-rank mixed states were computed\cite{tangle4,tangle5}. 

%%%%%%%%%%%%%%%%%%%%%%%%%%%%%%%%%%%%%%%%%%%%%%%%%%%%%%%%%
\begin{figure}[ht!]
\begin{center}
\includegraphics[height=5.3cm]{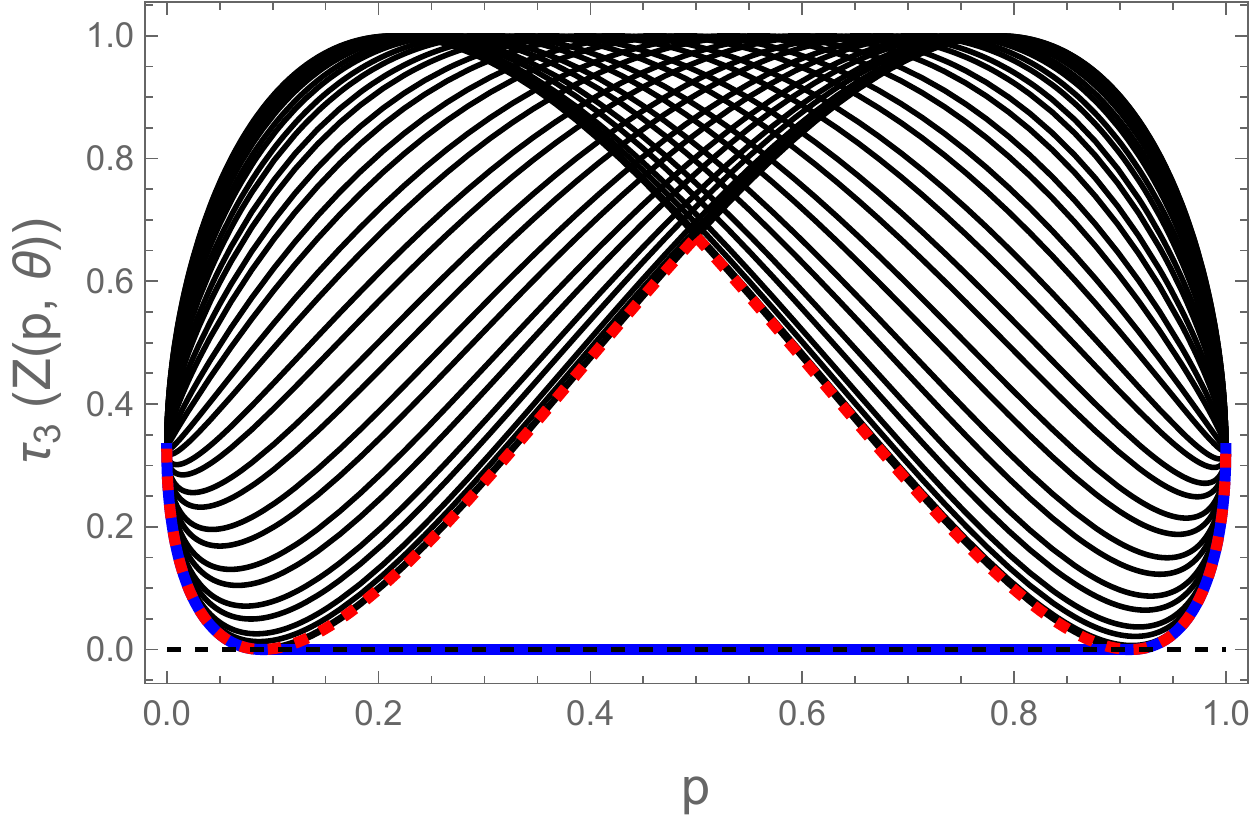} 
\includegraphics[height=5.3cm]{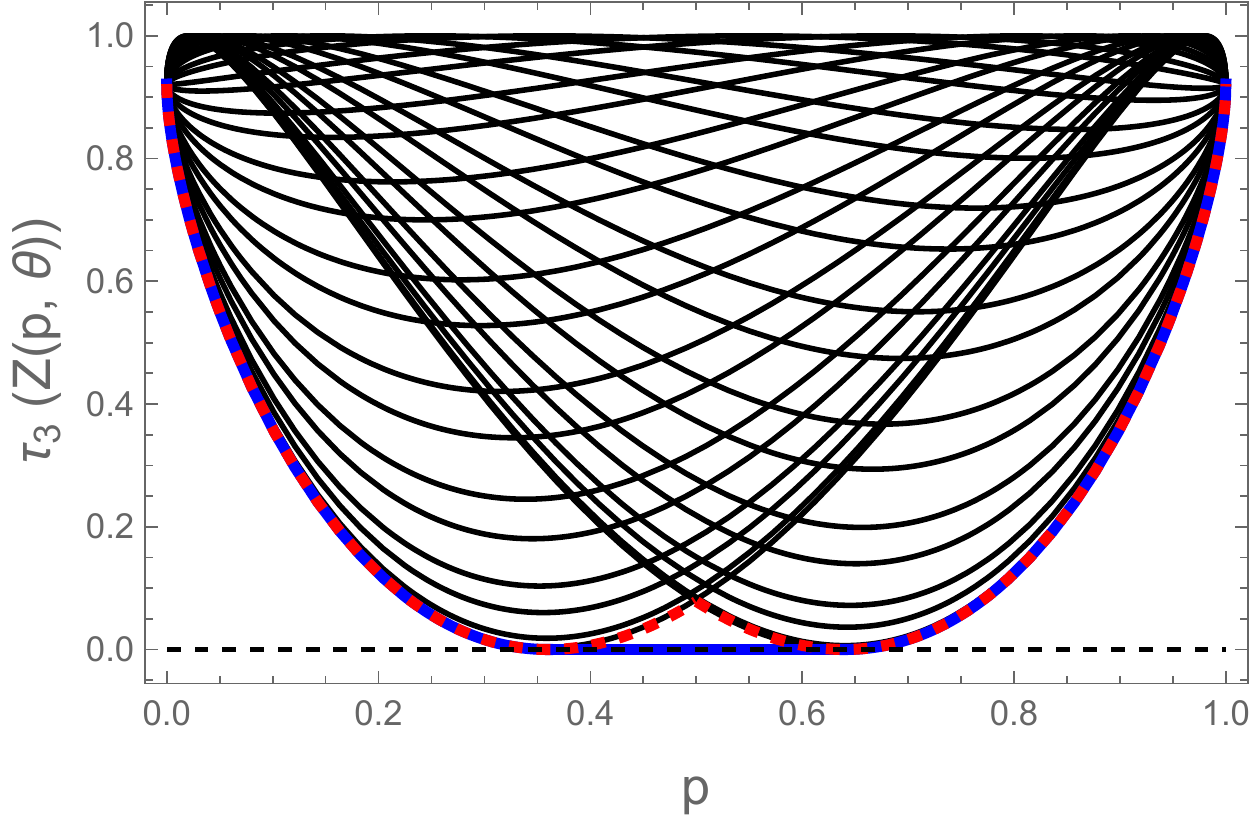}

\caption[fig2]{(Color online) Characteristic curves at (a) $x_1 = 0.3$ and (b) $x_1 = 0.8$. Minimum of the curves is plotted as a thick dashed (red) curve. The convex hull is plotted as a think blue curve.
 }
\end{center}
\end{figure}
%%%%%%%%%%%%%%%%%%%%%%%%%%%%%%%%%%%%%%%%%%%%%%%%%%%%%%%%%%%
In order to compute the three-tangle of $\rho$ for arbitrary $p$ we define
\begin{equation}
\label{rank2-4}
\ket{Z (p, \theta)} = \sqrt{p} \ket{\phi_1} - e^{i \theta} \sqrt{1 - p} \ket{\phi_2} = y_1 \left( \ket{010} - \ket{011} \right) + y_2 \left(\ket{100} + \ket{101} \right)
\end{equation}
where 
\begin{equation}
\label{rank2-5}
y_1 = \frac{1}{\sqrt{2}} \left[ \sqrt{p} x_1 + e^{i \theta} \sqrt{1 - p} x_2 \right]   \hspace{.5cm} y_2 = \frac{1}{\sqrt{2}} \left[ \sqrt{p} x_2 - e^{i \theta} \sqrt{1 - p} x_1 \right].
\end{equation} 
Then, one can show that the three-tangle for $\ket{Z (p, \theta)}$ is given by 
\begin{eqnarray}
\label{rank2-6}
\tau_3 \left(Z (p, \theta) \right) &=& 16 |y_1 y_2|^2                                                          \\     \nonumber
&=& 4 \Bigg[ \left[ p^2 + (1 - p)^2 - 2 p (1 - p) \cos 2 \theta \right] x_1^2 x_2^2                     \\     \nonumber
&& \hspace{1.0cm}                             + p (1 - p) (x_1^2 - x_2^2)^2 - 2 \sqrt{p (1 - p)} (2 p - 1) x_1 x_2 (x_1^2 - x_2^2) \cos \theta \Bigg].
\end{eqnarray}
The $\tau_3 \left(Z (p, \theta) \right)$ is plotted in Fig. 2 when (a) $x_1 = 0.3$ and (b) $x_1 = 0.8$ with choosing $\theta$ from $0$ to $2 \pi$ as an interval $0.2$.
Both figures show that $\tau_3 \left(Z (p, \theta) \right)$ is minimized by 
\begin{equation}
\label{rank2-7}
f(p) = \min \bigg[\tau_3 \left(Z (p, 0) \right), \tau_3 \left(Z (p, \pi) \right) \bigg].
\end{equation}
This curve is plotted in Fig. 2 as  thick dashed red lines.
Fig. 2 also show that $f(p) = 0$ at $p = p_{\pm}$, where
\begin{equation}
\label{rank2-8}
p_{\pm} = \frac{1}{2} \left[ 1 \pm |x_1^2 - x_2^2| \right].
\end{equation}
One can show that $f''(p)$ is negative in some region depending on $x_1$ and $x_2$ between $p_-$ and $p_+$. 
Therefore, $\tau_3 (\rho)$, the convex hull of $f(p)$, can be written in a form 
\begin{eqnarray}
\label{rank2-9}
\tau_3 (\rho) = \left\{                \begin{array}{cc}
                                        f(p)  & \hspace{.5cm}  p \leq p_-  \hspace{.2cm}  \mbox{or} \hspace{.2cm}  p \geq p_+                 \\
                                        0     &   p_- \leq p \leq p_+.
                                                   \end{array}                                  \right.        
\end{eqnarray}
This is plotted in Fig. 2 as thick blue line.

\section{Brief Review of HHL algorithm}

%%%%%%%%%%%%%%%%%%%%%%%%%%%%%%%%%%%%%%%%%%%%%%%%%%%%%%%%%
\begin{figure}[ht!]
\begin{center}
\includegraphics[height=6.0cm]{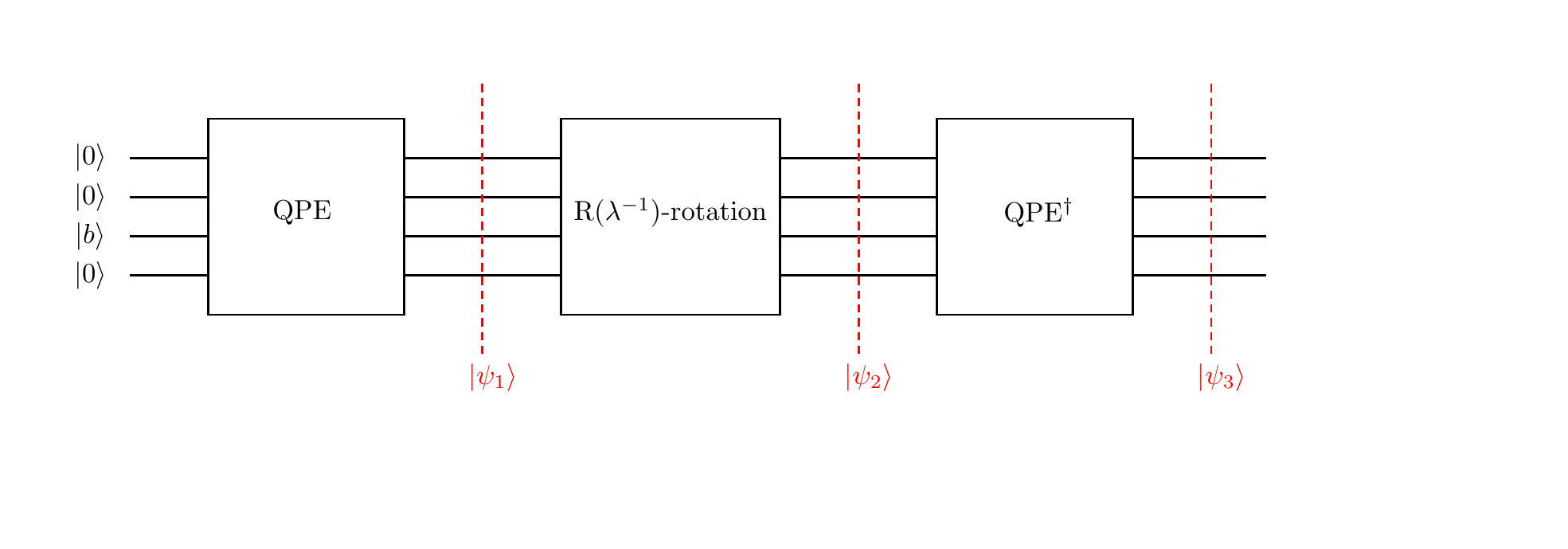} 

\caption[fig3]{(Color online) The schematic representation of the HHL algorithm.
}
\end{center}
\end{figure}
%%%%%%%%%%%%%%%%%%%%%%%%%%%%%%%%%%%%%%%%%%%%%%%%%%%%%%%%%%%

The HHL algorithm\cite{hhl} consists of three steps, which are QPE\cite{QPE1,QPE2,QPE3}, $R(\lambda^{-1})$-rotation, and inverse QPE. 
The schematic quantum circuit of the HHL algorithm is plotted in Fig. 3.
These three steps were experimentally and explicitly realized in Ref. \cite{exp-hhl-1}\footnote{ In Fig. 1 of Ref. \cite{exp-hhl-1} the rotations $R_y (\pi / 2^r)$ and $R_y (2 \pi / 2^r)$ should be interchanged.} by selecting a linear equation $A {\bm x} = {\bm b}$, where 
\begin{eqnarray}
\label{hhl-1}
A = \frac{1}{2}  \left(      \begin{array}{cc}
                                        3  &  1                   \\
                                        1  &  3 
                                        \end{array}           \right)          \hspace{1.0cm}
{\bm b} = \left(              \begin{array}{c}
                                       b_0   \\   b_1
                                       \end{array}           \right)
\end{eqnarray}
where $b_0^2 + b_1^2 = 1$. If $A$ is not hermitian, one can change the linear equation by $\tilde{A} {\bm y} = \left(   \begin{array}{c} {\bm b} \\  {\bm 0}  \end{array}  \right)$ where 
$\tilde{A}$ is a hermitian matrix given by $
A =   \left(      \begin{array}{cc}  0  &  A                   \\   A^{\dagger}  &  0  \end{array}           \right) $.

At the QPE stage we perform the QPE-algorithm by applying the unitary operator $e^{i A t}$ to $\ket{b} = b_0 \ket{0} + b_1 \ket{1}$. As shown in the appendix A, after the QPE stage the quantum state becomes 
\begin{equation}
\label{hhl-2}
\ket{\psi_1} = \sum_i \beta_i \ket{\lambda_i} \ket{u_i} \otimes \ket{0}
\end{equation}
if one chooses $t = 2 \pi / 2^n$. In Eq. (\ref{hhl-2}) $\lambda_i$ and $\ket{u_i}$ are the eigenvalue and corresponding eigenvector of $A$ and $\beta_i$ is defined by $\ket{b} = \sum_i \beta_i \ket{u_i}$. The last qubit $\ket{0}$ is an ancilla, which will be 
used at the next stage. For the case of Eq. (\ref{hhl-1}) we should choose $t = 2 \pi / 4$ and 
\begin{eqnarray}
\label{hhl-3}
&&\lambda_1 = 1 = (0 \hspace{.2cm} 1)_2       \hspace{1.0cm}   \ket{u_1} = \frac{1}{\sqrt{2}} \left( \ket{0} - \ket{1}  \right)               \\     \nonumber
&&\lambda_2 = 2 = (1 \hspace{.2cm} 0)_2       \hspace{1.0cm}   \ket{u_2} = \frac{1}{\sqrt{2}} \left( \ket{0} + \ket{1}  \right)               \\     \nonumber
&&  \beta_1 = \frac{1}{\sqrt{2}} (b_0 - b_1)  \hspace{1.0cm}    \beta_2 = \frac{1}{\sqrt{2}} (b_0 + b_1).     
\end{eqnarray}  
Thus, if the QPE stage is perfectly implemented in quantum computer, the quantum state $\ket{\psi_1}$ reduces to 
\begin{equation} 
\label{hhl-4}
\ket{\psi_1} = \frac{1}{2} \bigg[ (b_0 - b_1) \ket{01} \otimes (\ket{0} - \ket{1}) + (b_0 + b_1) \ket{10} \otimes  (\ket{0} + \ket{1}) \bigg] \otimes \ket{0}.
\end{equation}

In $R(\lambda^{-1})$-rotation stage we implement the controlled rotations to the ancilla qubit, whose angles are proportional to $\lambda_i^{-1}$. 
Using $\sin \theta \approx \theta$, the quantum state becomes approximately
\begin{equation}
\label{hhl-5}
\ket{\psi_2} \approx \sum_i \beta_i \ket{\lambda_i} \ket{u_i} \otimes \left[ \sqrt{1 - \frac{C^2}{\lambda_i^2}} \ket{0} + \frac{C}{\lambda_i} \ket{1} \right]
\end{equation}
after this stage, where $C$ is some appropriate nonzero constant.  For the case of Eq. (\ref{hhl-1}) $C$ is chosen as $C = (\sin \pi / 4 + 2 \sin \pi / 8) / 2 \approx 0.736$. 
If, therefore, this stage is perfectly implemented, after this stage $\ket{\psi_2}$ becomes approximately
\begin{equation}
\label{hhl-6}
\ket{\psi_2} \approx \frac{1}{2} \bigg[ (b_0 - b_1) \ket{01} \otimes (\ket{0} - \ket{1}) + (b_0 + b_1) \ket{10}  \otimes  (\ket{0} + \ket{1})  \bigg] \otimes \left[ \sqrt{1 - \frac{C^2}{\lambda_i^2}} \ket{0} + \frac{C}{\lambda_i} \ket{1} \right].
\end{equation}

At the $\mbox{QPE}^{\dagger}$ stage we undo the QPE implementation to uncompute the $\ket{\lambda_i}$. As a result, after this stage $\ket{\lambda_i}$ in Eq. (\ref{hhl-5}) is changed into $\ket{0}^{\otimes n}$, which gives 
\begin{equation}
\label{hhl-7}
\ket{\psi_3} = \ket{0}^{\otimes n} \otimes \sum_i \beta_i \ket{u_i} \otimes \left[ \sqrt{1 - \frac{C^2}{\lambda_i^2}} \ket{0} + \frac{C}{\lambda_i} \ket{1} \right].
\end{equation}
For the case of Eq. (\ref{hhl-1}) $\ket{\psi_3}$ reduces to 
\begin{eqnarray}
\label{hhl-8}
&&\ket{\psi_3} = \ket{00} \otimes \Bigg[ \frac{1}{2} \left\{ (b_0 - b_1) \sqrt{1 - C^2} + (b_0 + b_1) \sqrt{1 - \frac{C^2}{4}} \right\} \ket{00}                     \\    \nonumber
 &&       \hspace{3.5cm}                                                     +  \frac{1}{2} \left\{ - (b_0 - b_1) \sqrt{1 - C^2} + (b_0 + b_1) \sqrt{1 - \frac{C^2}{4}} \right\} \ket{10}                \\     \nonumber
&&         \hspace{5.0cm}                                                    + C (x_0 \ket{0} + x_1 \ket{1}) \otimes \ket{1}  \Bigg]
\end{eqnarray}
where 
\begin{eqnarray}
\label{hhl-9}
{\bm x} = A^{-1} {\bm b} = \frac{1}{4} \left(       \begin{array}{c}
                                                                       3 b_0 - b_1   \\  -b_0 + 3 b_1
                                                                            \end{array}                     \right)   \equiv \left(     \begin{array}{c}  x_0  \\   x_1    \end{array}  \right).
\end{eqnarray}
Therefore, measuring the ancilla qubit, one can compute $A^{-1} {\bm b}$ if the measurement outcome is $1$. 

\section{tripartite entanglement in HHL algorithm}

In this section we discuss how much the HHL algorithm utilizes the entanglement efficiently as we discussed previously in the Grover's algorithm. 
In this reason we will compute the entanglement at each stage of the HHL algorithm. For simplicity, we will consider only the case of Eq. (\ref{hhl-1}).
Thus, the three-tangle\cite{ckw} and $\pi$-tangle\cite{ou07-1} will be computed explicitly after taking a partial trace over the ancilla qubit in Eqs. (\ref{hhl-4}), (\ref{hhl-6}), and (\ref{hhl-8}). 

Since the ancilla qubit is decoupled in $\ket{\psi_1}$ of Eq. (\ref{hhl-4}), the first three-qubit state after the QPE stage is simply
\begin{equation}
\label{tri-1}
\ket{\bar{\psi}_1} = \frac{1}{2} \bigg[ (b_0 - b_1) \ket{01} \otimes (\ket{0} - \ket{1}) + (b_0 + b_1) \ket{10} \otimes  (\ket{0} + \ket{1}) \bigg].
\end{equation}
From Eq. (\ref{hhl-6}) one can compute $\bar{\rho}_2 = \mbox{Tr}_{\mbox{ancilla}} \ket{\psi_2} \bra{\psi_2}$. It turns out that $\bar{\rho}_2$ is rank-$2$ mixed state and its spectral decomposition is 
\begin{equation}
\label{tri-2}
\bar{\rho}_2 = p \ket{\phi_1} \bra{\phi_1} + (1 - p) \ket{\phi_2} \bra{\phi_2}
\end{equation}
where 
\begin{equation}
\label{tri-3}
p = \frac{1}{2} \left[ 1 + \sqrt{1 - 4 \beta_1^2 \beta_2^2 (1 - \gamma^2)} \right]   \hspace{1.0cm}      \gamma = \sqrt{\left(1 - C^2 \right) \left(1 - \frac{C^2}{4} \right)} + \frac{C^2}{2}.
\end{equation}
In Eq. (\ref{tri-2}) $\ket{\phi_1}$ and $\ket{\phi_2}$ are the same with Eq. (\ref{rank2-2}) if 
\begin{equation}
\label{tri-4}
x_1 = \frac{a_1}{\sqrt{a_1^2 + a_2^2}}     \hspace{1.0cm}   x_2 = \frac{a_2}{\sqrt{a_1^2 + a_2^2}} 
\end{equation}
where 
\begin{eqnarray}
\label{tri-5}
&&  a_1 = \beta_1 \left[ 1 + \sqrt{1 - 4 \beta_1^2 \beta_2^2 (1 - \gamma^2)} - 2 \beta_2^2 (1 - \gamma^2)  \right]       \\   \nonumber
&&  a_2 = \beta_2 \gamma \left[ 1 + \sqrt{1 - 4 \beta_1^2 \beta_2^2 (1 - \gamma^2)} \right].
\end{eqnarray}
Similarly, $\bar{\rho}_3 = \mbox{Tr}_{\mbox{ancilla}} \ket{\psi_3} \bra{\psi_3}$ can be computed by making use of Eq. (\ref{hhl-8}), whose spectral decomposition is 
\begin{equation}
\label{tri-6}
\bar{\rho}_3 = q \ket{\varphi_1} \bra{\varphi_1} + (1 - q) \ket{\varphi_2} \bra{\varphi_2}
\end{equation}
where 
\begin{equation}
\label{tri-7}
q = \frac{1}{2} \left[ 1 + \sqrt{1 - 4 (A C_2 - B C_1)^2} \right]
\end{equation}
with 
\begin{eqnarray}
\label{tri-8}
&& A = \frac{1}{2} \left[ (b_0 - b_1) \sqrt{1 - C^2} + (b_0 + b_1) \sqrt{1 - \frac{C^2}{4}} \right]                   \\    \nonumber
&& B = \frac{1}{2} \left[- (b_0 - b_1) \sqrt{1 - C^2} + (b_0 + b_1) \sqrt{1 - \frac{C^2}{4}} \right]                   \\    \nonumber
&& \hspace{1.0cm}  C_1 = C \frac{3 b_0 - b_1}{4}          \hspace{1.5cm}  C_2 = C \frac{-b_0 + 3 b_1}{4}.
\end{eqnarray}
One can show $A^2 + B^2 + C_1^2 + C_2^2 = b_0^2 + b_1^2 = 1$ explicitly. In Eq. (\ref{tri-6}) $\ket{\varphi_1}$ and $\ket{\varphi_2}$ are 
\begin{equation}
\label{tri-9}
\ket{\varphi_1} = \ket{00} \otimes \left( y_1 \ket{0} + y_2 \ket{1} \right)           \hspace{.5cm}
\ket{\varphi_2} = \ket{00} \otimes \left(- y_2 \ket{0} + y_1 \ket{1} \right)
\end{equation}
where 
\begin{equation}
\label{tri-10}
y_1 = \frac{f_1}{\sqrt{f_1^2 + f_2^2}}            \hspace{1.0cm}         y_2 = \frac{f_2}{\sqrt{f_1^2 + f_2^2}}  
\end{equation}
with
\begin{equation}
\label{tri-11}
f_1 = A^2 - B^2 + C_1^2 - C_2^2 + \sqrt{1 - 4 (A C_2 - B C_1)^2}      \hspace{1.0cm}  f_2 = 2 (A B + C_1 C_2).
\end{equation}

\subsection{Three-tangle}
Using Ref. \cite{ckw} it is easy to compute the three-tangle of $\ket{\bar{\psi}_1}$:
\begin{equation}
\label{tangle-1}
\tau_3 (\bar{\psi}_1) = 4 \beta_1^2 \beta_2^2 = (b_0^2 - b_1^2)^2.
\end{equation}
From Eq. (\ref{rank2-8}) $p_{\pm}$ for $\bar{\rho}_2$ is 
\begin{equation}
\label{tangle-2}
p_{\pm} = \frac{a_{\pm}^2}{a_1^2 + a_2^2}
\end{equation}
where $a_+^2 = \max (a_1^2, a_2^2)$ and $a_-^2 = \min (a_1^2, a_2^2)$.
Thus, making use of Eq. (\ref{rank2-9}) one can compute the three-tangle of $\bar{\rho}_2$, whose explicit expression is  
\begin{eqnarray}
\label{tangle-3}
\tau_3 (\bar{\rho}_2) = \left\{                \begin{array}{cc}
                                        g(p)  & \hspace{.5cm}  p \leq p_-  \hspace{.2cm}  \mbox{or} \hspace{.2cm}  p \geq p_+                 \\
                                        0     &   p_- \leq p \leq p_+.
                                                   \end{array}                                  \right.        
\end{eqnarray}
where $g(p) = \min \left[ g_+ (p), g_- (p) \right]$ with 
\begin{equation}
\label{tangle-4}
g_{\pm} (p) = \frac{4}{(a_1^2 + a_2^2)^2} \left[ (2 p - 1) a_1 a_2 \pm \sqrt{p (1 - p)} (a_1^2 - a_2^2) \right]^2.
\end{equation}
Of course, $p$, $a_1$, and $a_2$ are given in Eqs. (\ref{tri-3}) and (\ref{tri-5}). 
Since $\ket{\varphi_1}$ and $\ket{\varphi_2}$ are fully separable states, the three-tangle of $\bar{\rho}_3$ is 
\begin{equation}
\label{tangle-5}
\tau_3 (\bar{\rho}_3) = 0.
\end{equation}

%%%%%%%%%%%%%%%%%%%%%%%%%%%%%%%%%%%%%%%%%%%%%%%%%%%%%%%%%
\begin{figure}[ht!]
\begin{center}
\includegraphics[height=5.4cm]{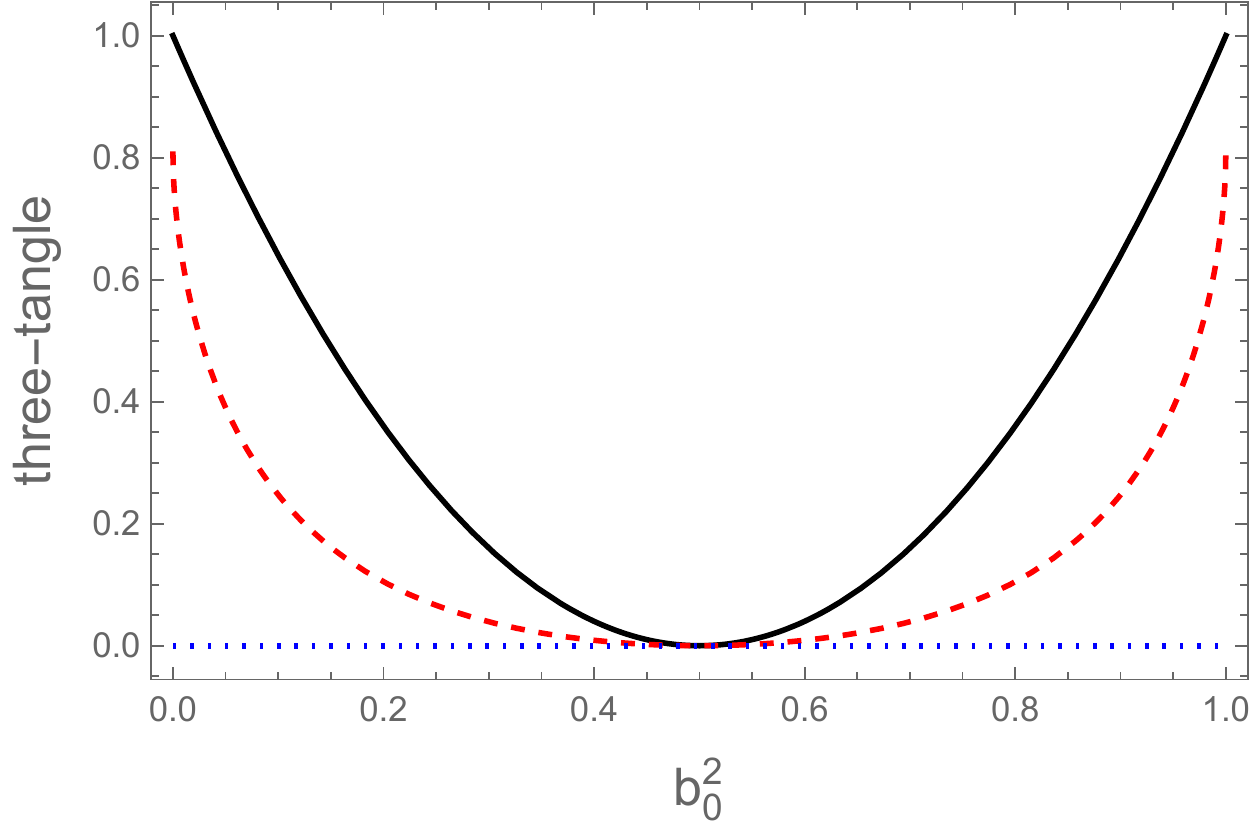} 
\includegraphics[height=5.4cm]{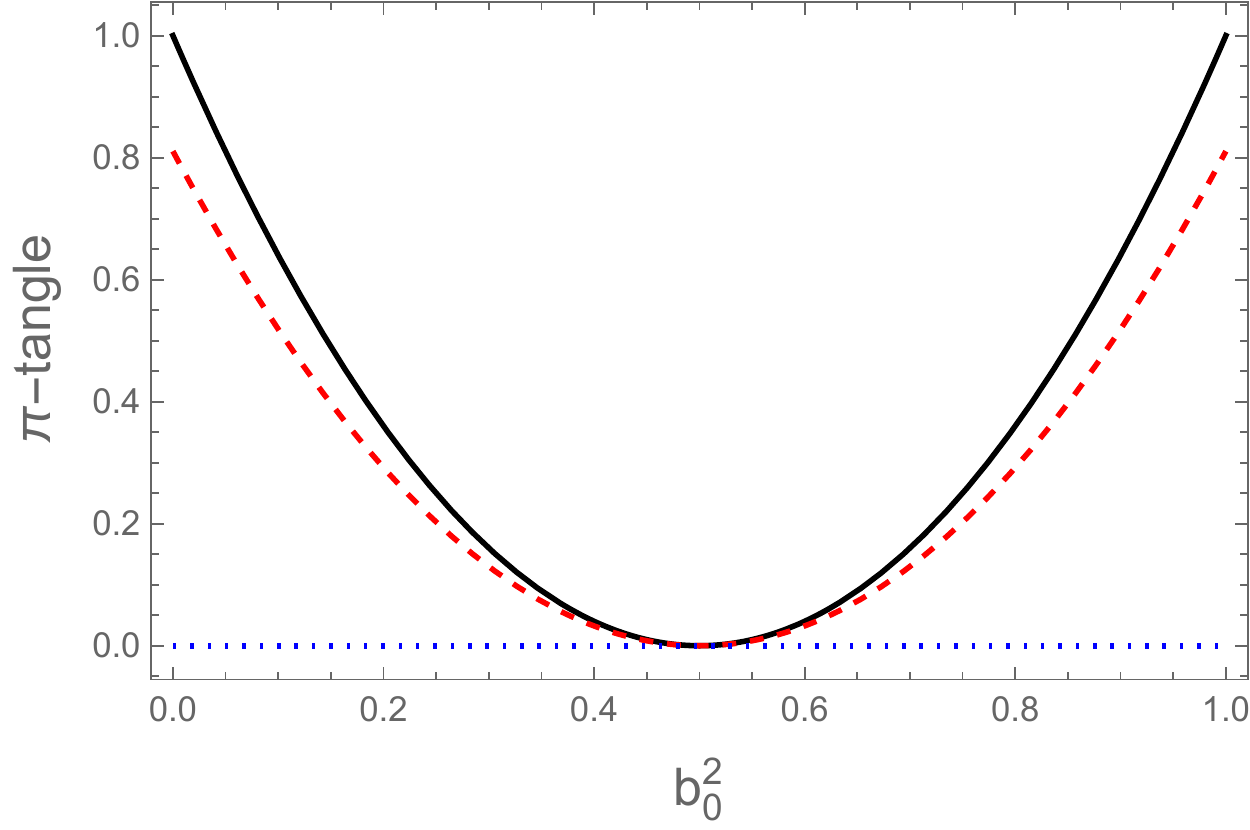}

\caption[fig4]{(Color online) (a) Plot of the three-tangles as a function of $b_0^2$. The black solid, red dashed, and blue dotted lines correspond to 
$\tau_3 (\bar{\psi}_1)$, $\tau_3 (\bar{\rho}_2)$, and $\tau_3 (\bar{\rho}_3)$. (b) Plot of the $\pi$-tangles as a function of $b_0^2$. The black solid, red dashed, and blue dotted lines correspond to $\pi_3 (\bar{\psi}_1)$, $\pi_3 (\bar{\rho}_2)$, and $\pi_3 (\bar{\rho}_3)$.
}
\end{center}
\end{figure}
%%%%%%%%%%%%%%%%%%%%%%%%%%%%%%%%%%%%%%%%%%%%%%%%%%%%%%%%%%%

The three-tangles for $\ket{\bar{\psi}_1}$, $\bar{\rho}_2$, and $\bar{\rho}_3$ are plotted in Fig. 4(a) as a function of $b_0^2$. 
The black solid, red dashed, and blue dotted lines correspond to 
$\tau_3 (\bar{\psi}_1)$, $\tau_3 (\bar{\rho}_2)$, and $\tau_3 (\bar{\rho}_3)$.
From this figure we understand that the QPE stage generates the three-tangle represented by the black solid line. This three-tangle reduces to red dashed line in
the $R(\lambda^{-1})$-rotation stage. Finally, the matrix inversion task is accomplished in the $\mbox{QPE}^{\dagger}$ stage at the price of  complete 
annihilation of the three-tangle. 

\subsection{$\pi$-tangle}
The $\pi$-tangle is a global negativity\cite{vidal01-1}-based tripartite entanglement measure\cite{ou07-1}. While three-tangle cannot detect the tripartite entanglement 
of the W-state class, $\pi$-tangle can detect it. 
For a three-qubit state $\rho$
the global negativities are given by
\begin{equation}
\label{negativity-1}
{\cal N}^A = || \rho^{T_A} || - 1, \hspace{1.0cm}
{\cal N}^B = || \rho^{T_B} || - 1, \hspace{1.0cm}
{\cal N}^C = || \rho^{T_C} || - 1,
\end{equation}
where $||R|| = \mbox{Tr} \sqrt{R R^{\dagger}}$, and the superscripts $T_A$, $T_B$ and $T_C$
represent the partial transposes of $\rho$ with respect to the qubits $A$, $B$ and $C$ respectively.
Using the separability criterion based on partial transpose~\cite{peres96,horod96,horod97},
it is easy to show that the global negativities vanish for separable states.
It is worthwhile noting that the computation of the global negativities is relatively simple
compared to the concurrence or three-tangle for mixed states since it does not need the convex-roof
extension defined in Eq. (\ref{mixed-3-tangle}). In addition, the negativities also satisfy the monogamy inequality
\begin{equation}
\label{monogamy-2}
{\cal N}_{AB}^2 + {\cal N}_{AC}^2 \leq {\cal N}_{A(BC)}^2
\end{equation}
like concurrence. Then, the $\pi$-tangle is defined as
\begin{equation}
\label{pi-1}
\pi_{ABC} = \frac{1}{3} (\pi_A + \pi_B + \pi_C ),
\end{equation}
where
\begin{equation}
\label{pi-2}
\pi_A = {\cal N}_{A(BC)}^2 - ({\cal N}_{AB}^2 + {\cal N}_{AC}^2) \hspace{.3cm}
\pi_B = {\cal N}_{B(AC)}^2 - ({\cal N}_{AB}^2 + {\cal N}_{BC}^2) \hspace{.3cm}
\pi_C = {\cal N}_{(AB)C}^2 - ({\cal N}_{AC}^2 + {\cal N}_{BC}^2).
\end{equation}
It is easy to show that the $\pi$-tangles for $|GHZ\rangle$ and $|W\rangle$ become
\begin{equation}
\label{pi-ghz-w}
\pi_{ABC} (|GHZ\rangle) = 1 \hspace{1.0cm}
\pi_{ABC} (|W\rangle) = \frac{4}{9} (\sqrt{5} - 1) \sim 0.55
\end{equation}
where 
\begin{equation}
\label{ghz-w}
\ket{GHZ} = \frac{1}{\sqrt{2}} (\ket{000} + \ket{111})               \hspace{1.0cm}
\ket{W} = \frac{1}{\sqrt{3}} (\ket{001} + \ket{010} + \ket{100} ).
\end{equation}
Thus, the $\pi$-tangle detects W-like(as well as GHZ-like) entanglement.

It is straightforward to compute the $\pi$-tangle of $\ket{\bar{\psi}_1}$, $\bar{\rho}_2$, and $\bar{\rho}_3$. The final result is 
\begin{equation}
\label{pi-tangle-1}
\pi_3 (\bar{\psi}_1) = 4 \beta_1^2 \beta_2^2 = (b_0^2 - b_1^2)^2               \hspace{.5cm}
\pi_3 (\bar{\rho}_2) = \frac{4 a_1^2 a_2^2}{(a_1^2 + a_2^2)^2} (2 p - 1)^2       \hspace{.5cm}
\pi_3 (\bar{\rho}_3) = 0.
\end{equation}
This is plotted in Fig. 4(b) as a function of $b_0^2$. 
The black solid, red dashed, and blue dotted lines correspond to 
$\pi_3 (\bar{\psi}_1)$, $\pi_3 (\bar{\rho}_2)$, and $\pi_3 (\bar{\rho}_3)$.
The behavior of the $\pi$-tangle is similar to that of the three-tangle. The only difference is that $\pi_3 (\bar{\psi}_1) - \pi_3 (\bar{\rho}_2)$ is generally 
smaller than $\tau_3 (\bar{\psi}_1) - \tau_3 (\bar{\rho}_2)$. This means that the decrease of $\pi$-tangle  in the  $R(\lambda^{-1})$-rotation stage is 
smaller compared to the three-tangle.

\section{Conclusion}
In this paper we examine a question `how much the HHL algorithm exploits the quantum entanglement efficiently?'.
In this reason we computed the tripartite entanglement explicitly at the every steps of the HHL algorithm by choosing the explicit example (\ref{hhl-1}). 

As Fig. 4 shows, the three-tangle and $\pi$-tangle exhibit similar behavior. At the QPE stage the tripartite entanglement is generated. 
The amount of it is dependent on $b_0^2$. If $b_0^2 = 0$ or $1$, maximal tripartite entanglement is generated at this stage. For other case partial entanglement is 
generated. At the $R(\lambda^{-1})$-rotation stage some entanglement is used. Thus, the net tripartite entanglement is reduced. 
At the final $\mbox{QPE}^{\dagger}$ stage the matrix inversion task is completed at the price of vanishing all entanglement. 

The HHL algorithm is known as an optimal\cite{hhl} in the matrix inversion task. It is also known that entanglement is a physical resource in  
quantum information processing. Then, the following question arises: why maximal creation of 
entanglement and complete annihilation of it do not occur? As discussed in Sec. II similar phenomenon happens in the Grover's algorithm for large $N$
even though it is known to be optimal\cite{zalka99} in the quantum searching task. This means that from an aspect of entanglement, both algorithms do not use 
entanglement efficiently to some extent. We think there are additional quantities as well as entanglement, which may play important role in the quantum algorithm.
Similar situation happens in the quantum illumination\cite{lloyd08,tan08,eylee-22-1}. In this case the quantum discord is known to be important\cite{discord1,discord2}. 

Even though we have not presented explicitly in the paper, one can show that all bipartite entanglement of $\ket{\bar{\psi}_1}$ , $\bar{\rho}_2$, and $\bar{\rho}_3$ are zero. This means that 
$\ket{\bar{\psi}_1}$ and $\bar{\rho}_2$ are in the GHZ-class\cite{dur00-1}. We do not exactly understand why the W-type entanglement does not 
arise. All the questions will be discussed in the future.

%{\bf Acknowledgement}:
%On April 16, 2014 the ferry Sewol has sunk into the South Sea of Korea. Due to this disaster 304 people died and, 9 of them are still missing. We would like to dedicate this paper to all victims of this accident.
%This research was supported by the Basic Science Research Program through the National Research Foundation of Korea(NRF) funded by the Ministry of Education, Science and Technology(2011-0011971).
%This work was supported by the Kyungnam University Foundation Grant, 2017.

\begin{figure}[ht!]
\begin{center}
\includegraphics[height=6.0cm]{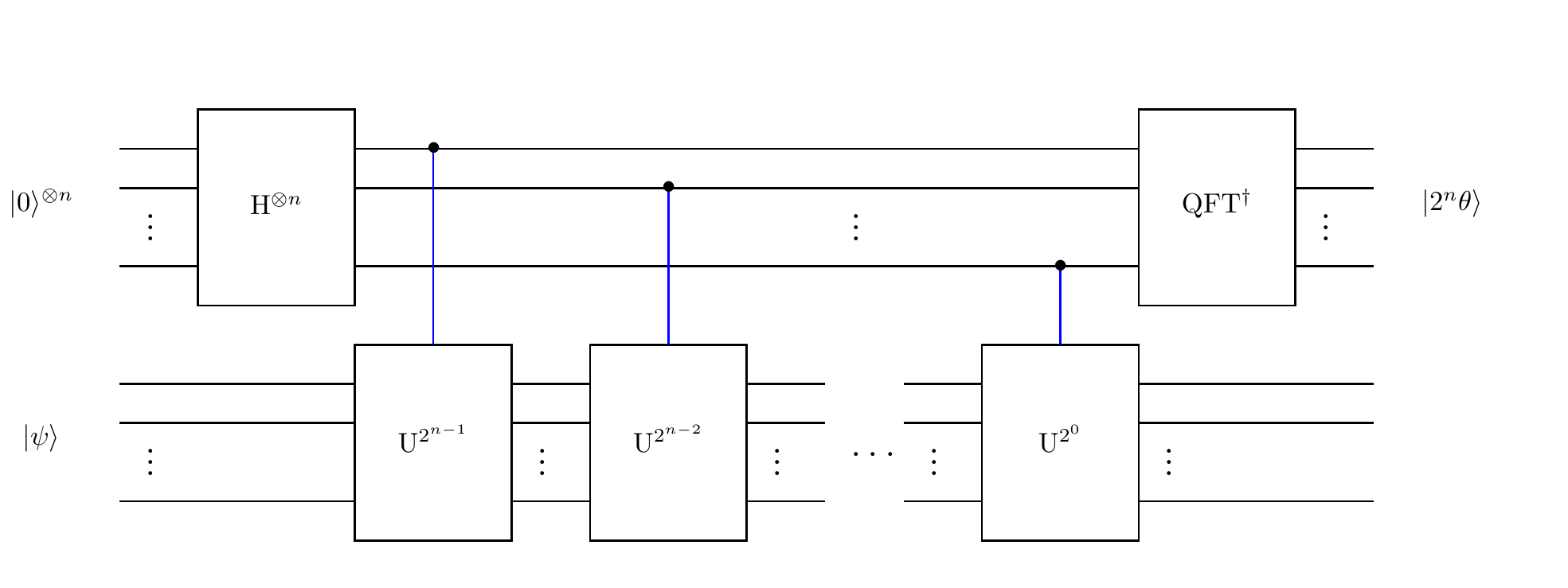} 

\caption[fig5]{(Color online) The quantum circuit of  QPE. In the figure QFT means quantum Fourier transform. If $\ket{\psi}$ is an eigenvector of the unitary operator $U$ in a form $U \ket{\psi} = e^{2 \pi i \theta} \ket{\psi}$, 
this QPE quantum circuit generates $\ket{2^n \theta}$ in the c-register. }
\end{center}
\end{figure}
%%%%%%%%%%%%%%%%%%%%%%%%%%%%%%%%%%%%%%%%%%%%%%%%%%%%%%%%%%%
In this appendix QPE will be discussed in detail and Eq. (\ref{hhl-2}) will be derived explicitly. The quantum circuit of QPE\cite{QPE3} is plotted in Fig. 5. 
In Fig. 5 QFT stands for quantum Fourier transform\cite{text,shor94,QFT1}, which is defined by the unitary operator $U_{QFT}$ obeying
\begin{equation}
\label{QFT-1}
\ket{\tilde{x}} = U_{QFT} \ket{x} = \frac{1}{\sqrt{N}} \sum_{y=0}^{N-1} e^{\frac{2 \pi i x \cdot y}{N}} \ket{y}
\end{equation}
where $\ket{x}$ is arbitrary $n$-qubit state and $N = 2^n$. As Fig. 5 shows, QPE consists of two registers. First register starts with $\ket{0}^{\otimes n}$, 
which we will call `c-register'. Second register starts with some quantum state $\ket{\psi}$, which we will call `t-register'. 
In the figure $U$ is a certain unitary operator.

If $\ket{\psi}$ is an eigenvector of $U$ in a form
\begin{equation}
\label{appen-1}
U \ket{\psi} = e^{2 \pi i \theta} \ket{\psi},
\end{equation}
the quantum circuit in Fig. 5 generates $\ket{2^n \theta} \otimes \ket{\psi}$ in the c-register and t-register respectively at the final stage. 
If the whole process of QPE is represented by an unitary operator $U_{QPE}$, this fact implies 
\begin{equation}
\label{appen-2}
U_{QPE} \bigg[ \ket{0}^{\otimes n} \otimes \ket{\psi} \bigg] = \ket{2^n \theta} \otimes \ket{\psi}.
\end{equation}
Thus, measuring the c-register at the final step, one can estimate the 
phase $\theta$. For example, if the outcome is $m$, one can conjecture $\theta = m / 2^n.$\footnote{If $\theta$ is not a form of $m / 2^n$, the quantum circuit
produces an inequality $\theta_{min} < \theta < \theta_{max}$, where $\theta_{max} - \theta_{min}$ decreases with increasing $n$.}

Now, let us consider the case that $\ket{\psi}$ is not an eigenvector of $U$. Let the eigenvectors of $U$ be $\ket{u_j}$ obeying
\begin{equation}
\label{appen-3}
U \ket{u_j} = e^{2 \pi i \theta_j} \ket{u_j}.
\end{equation}
Since a set $\{ \ket{u_j} \}$ is a complete, $\ket{\psi}$ can be expressed as a linear combination of $\ket{u_j}$ in a form
\begin{equation}
\label{appen-4}
\ket{\psi} = \sum_{j} \beta_j \ket{u_j}.
\end{equation}
Using a fact that $U_{QPE}$ is a linear operator, it is straightforward to show 
\begin{equation}
\label{appen-5}
U_{QPE} \bigg[ \ket{0}^{\otimes n} \otimes \ket{\psi} \bigg] =  \sum_{j} \beta_j \ket{2^n \theta_j} \otimes \ket{u_j}
\end{equation}
in this case. 

Since $U = e^{i A t}$ and $\ket{\psi} = \ket{b}$ in the HHL algorithm and they obey
\begin{equation}
\label{appen-6}
A \ket{u_j} = \lambda_j \ket{u_j}      \hspace{1.0cm}   \ket{b} = \sum_j \beta_j \ket{u_j},
\end{equation}
the final state after QPE becomes 
\begin{equation}
\label{appen-7}
\ket{\psi_{final}} = \sum_j \beta_j \ket{2^n \frac{\lambda_j}{2 \pi} t} \otimes \ket{u_j}.
\end{equation}
Thus, if $t$ is chosen as $t = 2 \pi / 2^n$, one can derive Eq. (\ref{hhl-2}).

\end{appendix}

\end{document}